\documentclass[useAMS,usenatbib]{mn2e}
\usepackage{graphicx}

\title[33 RR Lyrae stars observed with K2-E2]{An RR Lyrae family portrait: 33 stars observed in Pisces with K2-E2}
\author[L.\ Moln\'ar et al.]{L.\ Moln\'ar$^{1,2}$\thanks{E-mail: molnar.laszlo@csfk.mta.hu}, R.\ Szab\'o$^1$, P.\ A.\ Moskalik$^3$, J.\ M.\ Nemec$^4$, E.\ Guggenberger$^{5,6}$,
\and R.\ Smolec$^3$, R.\ Poleski$^7$, E.\ Plachy$^{1,2}$, K.\ Kolenberg$^{8,9,10}$, Z.\ Koll\'ath$^{1,2}$\\
$^{1}$Konkoly Observatory, MTA CSFK, Konkoly Thege Mik\'os \'ut 15-17, H-1121 Budapest, Hungary\\
$^{2}$Institute of Mathematics and Physics, Savaria Campus, University of West Hungary\\ K\'arolyi G\'asp\'ar t\'er 4, H-9700 Szombathely, Hungary\\
$^3$Copernicus Astronomical Center, Polish Academy of Sciences, ul.\ Bartycka 18, 00-716 Warszawa, Poland\\
$^4$Department of Physics \& Astronomy, Camosun College, Victoria, British Columbia, V8P 5J2, Canada\\
$^5$Max Planck Institut f\"ur Sonnensystemforschung, Justus-von-Liebig-Weg 3, 37077 G\"ottingen, Germany\\
$^6$Stellar Astrophysics Centre, Department of Physics and Astronomy, Aarhus University, DK-8000 Aarhus C, Denmark\\
$^7$Department of Astronomy, Ohio State University, 140 W.\ 18th Ave., Columbus, OH 43210, USA\\
$^8$Instituut voor Sterrenkunde, Celestijnenlaan 200D, B-3001 Leuven, Belgium\\
$^9$Harvard-Smithsonian Center for Astrophysics, 60, Garden street, Cambridge MA 02138, USA\\
$^{10}$Department of Physics, University of Antwerp, Groenenborgerlaan 171, 2020 Antwerp, Belgium\\
}
\begin{document}

\date{Accepted 1988 December 15. Received 1988 December 14; in original form 1988 October 11}

\pagerange{\pageref{firstpage}--\pageref{lastpage}} \pubyear{2014}

\maketitle

\label{firstpage}

\begin{abstract}
A detailed analysis is presented of 33 RR Lyrae stars in Pisces observed with the \textit{Kepler} space telescope over the 8.9-day long K2 Two-Wheel Concept Engineering Test. The sample includes not only fundamental-mode and first overtone (RRab and RRc) stars but the first two double-mode (RRd) stars that \textit{Kepler} detected and the only modulated first-overtone star ever observed from space so far. The precision of the extracted K2 light curves made it possible to detect low-amplitude additional modes in all subtypes. All RRd and non-modulated RRc stars show the additional mode at $P_X/P_1\sim 0.61$ that was detected in previous space-based photometric measurements. A periodicity longer than the fundamental mode was tentatively identified in one RRab star that might belong to a gravity mode. We determined the photometric [Fe/H] values for all fundamental-mode stars and provide the preliminary results of our efforts to fit the double-mode stars with non-linear hydrodynamic pulsation models. The results from this short test run indicate that the K2 mission will be, and has started to be, an ideal tool to expand our knowledge about RR Lyrae stars. As a by-product of the target search and analysis, we identified 165 \textit{bona-fide} double-mode RR Lyrae stars from the Catalina Sky Survey observations throughout the sky, 130 of which are new discoveries.
\end{abstract}


\section{Introduction}
RR Lyrae stars are pulsating low-mass horizontal-branch stars. Since they can be found at the intersection of the horizontal branch and the classical instability strip, a relatively small region of the Hertzsprung-Russell-diagram, their absolute brightnesses are fairly similar, making them good distance indicators. The large number of RR~Lyrae stars make them good tracers of halo structures around the Milky Way \citep{sesar} and even in nearby galaxies (e.g.\ \citealt{m31,m33,lmc}). The period ratios of the rare double-mode stars can be used to determine the masses of RR Lyrae stars independent from stellar evolution theory. 

The \textit{Kepler} space telescope observed $\sim45$ RR~Lyrae stars in the Lyra-Cygnus field of the prime mission, spanning a wide metallicity and brightness range, from the eponym of the class, RR~Lyr ($Kp = 7.862$ mag, \citealt{rrlyr}), down to about $Kp =17.5$ magnitudes \citep{nemec13}. Although the \textit{Kepler} mission was designed to detect subtle light changes caused by planetary transits \citep{borucki} and the high-amplitude variations of the RR~Lyrae stars required additional processing, space-based photometry proved to be essential to further our understanding of RR Lyrae stars \citep{benko14}. 

\textit{Kepler} observed mostly fundamental-mode stars, with only four first-overtone and no double-mode stars in the sample. Therefore most of its findings have been limited to the RRab class. The results include the discovery of period doubling \citep{kolenberg10,pd}, a wealth of low-amplitude additional modes \citep{benko10,gug12} and the discovery that most Blazhko stars show multiple or irregular modulations \citep{benko14}. The observations also revealed that the new phenomena (period doubling and additional modes) are strictly limited to the modulated stars \citep{nemec11}, and a recent analysis of stars observed by the \textit{CoRoT} space telescope confirmed this dichotomy \citep{szabo14}. In contrast, all four first-overtone stars observed by \textit{Kepler} (and two more in the \textit{CoRoT} sample) exhibit the same additional mode at the frequency ratio $f_1/f_2 \approx 0.60-0.64$, without any signs of modulation in the main mode \citep{moskalik_rrc,moskalik14}. The origin of this additional mode is uncertain. We note that we refer to this mode as $f_X$ instead of $f_2$ to avoid any confusion with the suspected second-overtone signals in the RRab stars.

The primary mission of \textit{Kepler} gathered four years of (quasi)-continuous observations. These allowed for more specific studies, such as following the very unexpected vanishing of the Blazhko effect in RR Lyr \citep{leborgne14} or the first attempt to search for chaos in the modulation of an RRab star \citep{plachy14}, and even the search for artificial origins in the period variations \citep{hippke14}. After the unfortunate failure of the second reaction wheel, however, maintaining the original pointing was not feasible any more. 

The new K2 mission of the \textit{Kepler} space telescope uses orientations along the Ecliptic where the spacecraft can balance against the radiation pressure of the Sun \citep{howell14}. This new strategy results in short, $\sim75$ day long observing campaigns that scan various fields in the sky, reaching targets the original field lacked. In this paper we detail the results from the very first measurements of the K2 mission. Section 2 describes the K2-E2 run and the extraction of the light curves. Results are detailed in Sections 3 and 4 for RRd-RRc and RRab stars, respectively. Conclusions and future directions are summarized in Section 5. 

\section{K2 data and reduction}
\textit{Kepler} observed a stellar field around the vernal equinox point in Pisces (center coordinates: $\alpha= 359\degr$, $\delta= -2\degr$) between 4th-13th February 2014. The primary goal of this K2 Two-Wheel Concept Engineering Test (hereafter K2-E2) was to test the performance of the telescope in fine guidance mode. As well, the observations of nearly 2000 targets were made available for the scientific community. Instead of the tight target pixel masks used in the primary mission, larger subframes were stored around the targets in order to avoid any flux loss due to the drift of the telescope. A $\sim 1$ pixel variation in the position of the stars was observed during the 8.9-day long K2-E2 data. The boresight was shifted by multiple pixels 2.3 days into the test, at BJD = 56695.36, when the telescope locked successfully into fine guidance mode. Fortunately the 50x50-pixel target masks were more than enough to contain these movements.

We identified 34 potential RR Lyrae stars in the K2-E2 sample and extracted their photometric data with the {\sc PyKE} software\footnote{\texttt{http://keplergo.arc.nasa.gov/PyKE.shtml}}, developed for the \textit{Kepler} mission by the \textit{Kepler} Guest Observer Office \citep{pyke}. All targets were observed in long cadence mode, with a sampling of 29.4 minutes. To counter the effect of drifting and capture all the flux, we defined relatively large pixel apertures for the stars, with 1--2-pixel-wide halos around the actual point spread functions. We applied two different masks for the positions before and after the boreshight shift. We used the background-corrected SAP (Simple Aperture Photometry) fluxes calculated by the {\sc kepextract} routine. The photometric precision spreads between $3.5\, \mu{\rm mag}$ to 7 mmag per data point for a 10.5 and a 17.4 mag star, respectively. In some cases we had to apply small amounts of shift and/or scaling to the data to connect the two parts of the light curves. These differences were likely caused by different pixel sensitivities in the two positions. Some stars are affected by the Fine Guidance Sensor clock crosstalk patterns \citep{ihandbook}. The faintest star in the sample, EPIC 60018780, suffers from extensive negative video crosstalk from a bright target elsewhere on the same module: dark CCD columns cross the image of the star. As the image of the star moves in and out of the dark column, jumps up to 0.1--0.2 mag appear in the light curve of this star, especially in the first 2.3 days.

Without the third reaction wheel, radiation pressure from the Sun slowly turns \textit{Kepler} away from the desired spacecraft attitude, and rolls it about the optical axis (the X-axis of the spacecraft, \citealt{howell14}). These disturbances are periodically corrected with the on-board thrusters. For some stars the frequency of the manoeuvres, $f_{\rm corr}=8.16\, {\mathrm d^{-1}}$, and its harmonics can be detected in the Fourier spectra. (We note that the K2-E2 run tested a 3-hour adjustment frequency, but later campaigns use 6-hour intervals, corresponding to $f_{\rm corr}=4.08\, {\mathrm d^{-1}}$.) Rolling of the field-of-view means that stars close to the center are least affected but targets close to the edges can experience considerable drifting. Although these peaks are well-defined and can be easily excluded from the analysis, they indicate that special care is needed to define the target apertures and to correct for pixel sensitivity variations.

\begin{table}
\caption{Sample of the complete K2-E2 light curve file. The full table is accessible online.} 
\begin{tabular}{lccc}
\hline
\noalign{\vskip 0.1cm}
EPIC   &   BJD-2450000   &   \textit{Kp}   &   $\Delta Kp$ \\
type: RRd   &   (d)  &   (mag)   &   (mag) \\
\hline
60018653   &   1860.0503350   &   13.90840   &   0.00054 \\
60018653   &   1860.0707686   &   13.92217   &   0.00054 \\
60018653   &   1860.0912023   &   13.92438   &   0.00054 \\
60018653   &   1860.1116358   &   13.91703   &   0.00054 \\
60018653   &   1860.1320694   &   13.89827   &   0.00053 \\
\multicolumn{4}{l}{\dots}\\
\hline
\end{tabular}
\label{lightcurve_table}
\end{table}
\begin{table*}
\caption{Sample of the frequency tables of the entire K2-E2 data set: frequencies, amplitudes, phases, the respective uncertainties and the identifications of the peaks ($f_0$: fundamental mode, $f_1$: first overtone, $f_X$, $f_2$, $f_g$: additional modes). Super-Nyquist frequency components are marked with asterisks. The entire table is accessible in the online version of the paper.}
\begin{tabular}{lccccccr}
\hline
\noalign{\vskip 0.1cm}
Star  & \textit{F} (d$^{-1}$) & \textit{A} (\textit{Kp} mag) & $\phi$ (rad/$2\pi$) & $\Delta F$ (d$^{-1}$) & $\Delta A$ (\textit{Kp} mag)  & $\Delta \phi$ (rad/$2\pi$ )  & ID \\  
\hline
60018653  & 1.85382  &  0.04641  & 0.92372  &  0.00031  & 0.00040  & 0.00080  & $f_0$\\
60018653  & 2.48564  &  0.13988  & 0.82381  &  0.00010  & 0.00035  & 0.00026  & $f_1$\\
60018653  & 3.7076    &  0.00422  & 0.2164    &  0.0034    & 0.00034  & 0.0088  & $2f_0$\\
60018653  & 5.561      &  0.00105  & 0.573      &  0.014      & 0.00037  & 0.035    & $3f_0$\\
60018653  & 4.97128  &  0.02273  & 0.1761    &  0.00063  & 0.00036  & 0.0016  & $2f_1$ \\
\multicolumn{8}{l}{\dots}\\
\hline
\end{tabular}
\label{freqtable}
\end{table*}

Undoubtedly, there will be better and more precise ways to extract K2 photometry as the mission evolves. However, (semi-)automated methods are usually optimised for relatively quiet stars and not for large-amplitude, short-period pulsators. One such pipeline, developed by \citet{vanderburg}, for example, admittedly failed to improve the light curves of RR Lyrae stars and eclipsing binaries. Therefore we were satisfied with the level of accuracy we achieved with the above method for this initial study. The light curves (along with the frequency tables, see Section \ref{sec_sample}) are available online for further analysis alongside the paper and at the \textit{Kepler} Investigations at Konkoly webpage\footnote{\texttt{http://konkoly.hu/KIK/data.html}}. Samples are provided in Tables \ref{lightcurve_table} and \ref{freqtable}.

\subsection{The K2-E2 RR Lyrae sample}
\label{sec_sample}
Prior to the K2-E2 test, a call was issued within KASC (the \textit{Kepler} Asteroseismic Science Consortium) to provide targets within the approximate position of the field in Pisces. We proposed 143 RR Lyrae stars, largely based on the RRab catalogue of the Catalina Sky Survey (CSS, \citealt{css,mls}) but also on the ASAS database and the SIMBAD identifications. By the time the K2-E2 data became available, the detailed classifications of periodic variables in CSS were published \citep{cssfull}, allowing for a follow-up search in the target masks. We found 34 stars at or close to the positions of the 1952 K2-E2 targets based on these catalogues. Comparison with the proposed list revealed that several of the originally proposed targets ended up off silicon, while others were not selected for observation.

One of the stars, EPIC 60018241 (ASAS J234439-0148.6) turned out to be an eclipsing binary \citep{conroyk2}, but the other 33 stars are RR Lyrae stars. Surprisingly, we identified RR Lyrae stars from all subclasses, including two classical double-mode (RRd) stars and a potential strongly modulated first-overtone (RRc) candidate. Basic parameters, i.e.\ the EPIC (\textit{Kepler} Ecliptic Plane Input Catalog) identification number, coordinates, brightness, period, other identifications, and subclass, are summarised in Table \ref{target_table} for RRd and RRc stars. The parameters of RRab stars are listed in \ref{rrab_target_table}. 

\begin{figure*}
\includegraphics[width=1.0\textwidth]{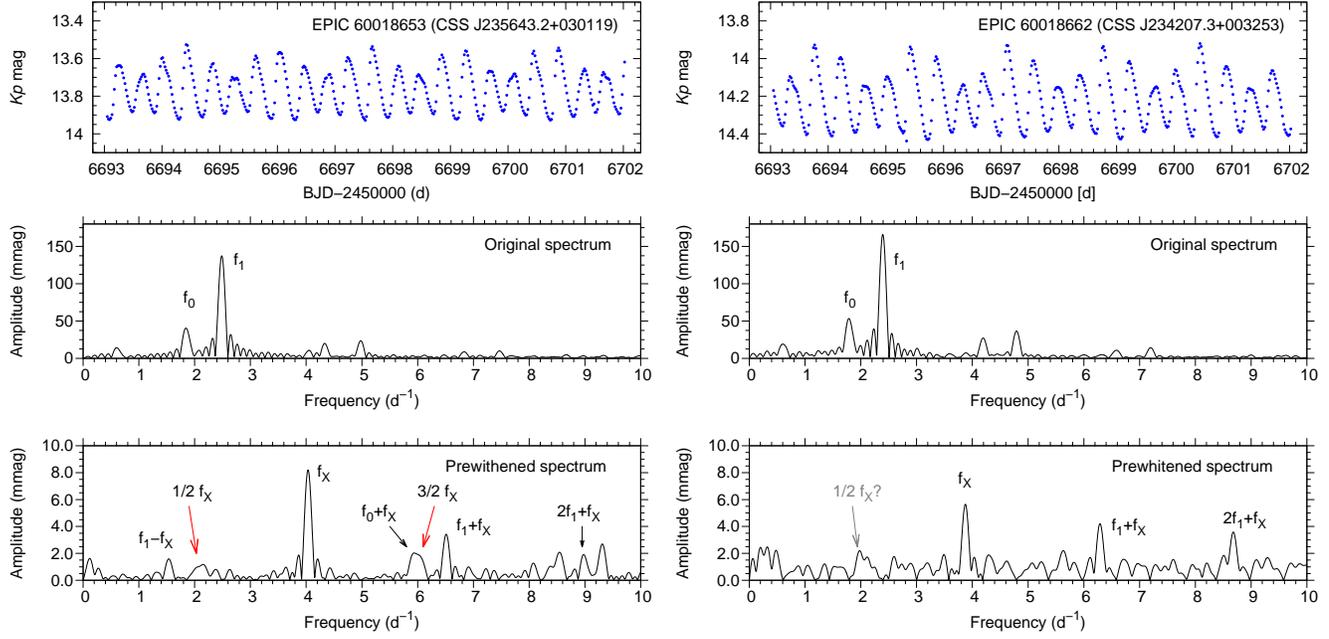}
\caption{Light curves (top row) and Fourier-spectra of the two RRd stars in the sample. Middle row: the original Fourier spectra. Fundamental-mode and first-overtone frequency peaks are labelled with $f_0$ and $f_1$. Bottom row: residual spectra after prewithening the data with $f_0$, $f_1$ and their linear combination peaks. The additional mode $f_X$ and its significant subharmonic peaks and combinations are labelled. We also indicated the marginally detected $1/2\,f_X$ peak in EPIC 60018662. }
\label{rrd}
\end{figure*}

We used the \texttt{Period04} software to calculate the Fourier transforms \citep{period04}. The frequency tables of the K2-E2 sample are summarised in Table \ref{freqtable}. Here we included only the resolved frequency peaks, i.e.\ unresolved residuals that indicate amplitude and/or phase variation are not listed. In some cases high-order harmonics reflect back from the Nyquist frequency ($f_N = 24.4684$ d$^{-1}$), but we fitted the super-Nyquist values in all cases. These are marked with asterisks in the frequency table.  

\begin{table*}
\caption{Double-mode and first-overtone RR Lyrae stars observed during the Two-Wheel Concept Engineering Test run of the K2 mission. O1 and FM indicate the periods of the first overtone and fundamental mode, respectively. We provide estimated \textit{Kp} magnitudes for two stars (in italics). The brightness of EPIC 60018238 is too high in the EPIC ($Kp=9.83$), compared to the flux values. The last star has no EPIC ID but was observed within the pixel mask of the engineering target EPIC 60042292. **Modulation period. }
\begin{tabular}{lcccclccc}
\hline
\noalign{\vskip 0.1cm}
EPIC ID     &   RA (deg)    &    Dec (deg)   &    \textit{Kp} mag &  O1 period (d)  & Name                &   FM period (d)  &    CSS ID number   \\
\hline
60018653   &    359.18015   &    +3.022110    &    13.760   &    0.402311   &    CSS J235643.2+030119   &    0.539427    &    1104128004573   \\ 
60018662   &    355.53058   &    +0.547972    &    14.216   &    0.417448   &    CSS J234207.3+003253   &    0.558996    &    1101127011298   \\ 
\hline
\noalign{\vskip 0.1cm}
60018224   &    358.53385   &    +0.963635    &    10.600   &    0.306156   &    EV Psc                 &    --    &    --   \\ 
60018238   &    354.87500   &    $-$3.631667   &    \textit{12.419}    &    0.274804   &    ASAS J233930$-$0337.9    &    --    &    --   \\ 
60018678   &    352.16375   &    +1.028500    &    14.862   &    0.432426   &    CSS J232839.2+010142   &    --    &    1101126021018   \\ 
(60042292)  &    359.42500   &    $-$1.839444   &    \textit{14.502}   &    0.300110   &    CSS J235742.1$-$015022   &    17.07**    &    1001128017680   \\ 
\hline
\end{tabular}
\label{target_table}
\end{table*}

\section{Double-mode and first-overtone stars}
\subsection{Double-mode stars}
We identified two RRd-type stars in the field. Both stars, EPIC 60018653 (CSS J235643.2+030119) and 60018662 (CSS J234207.3+003253), were misclassified in the Catalina sample, as RRc and RRab stars, respectively, but the K2 light curves revealed that they are actually classical double-mode RR Lyraes (Figure \ref{rrd}). These are the first RRd stars observed by \textit{Kepler} and only the third and fourth from space. AQ Leo, the brightest of the class, was observed by \textit{MOST} \citep{aqleo}, and the \textit{CoRoT} data of another star, CoRoT ID 0101368812, was analysed by \citet{chadid} and \citet{szabo14}. 

\subsubsection{Period ratios and physical properties}
\label{sec_hd_model}
\begin{figure}
\includegraphics[width=1.0\columnwidth]{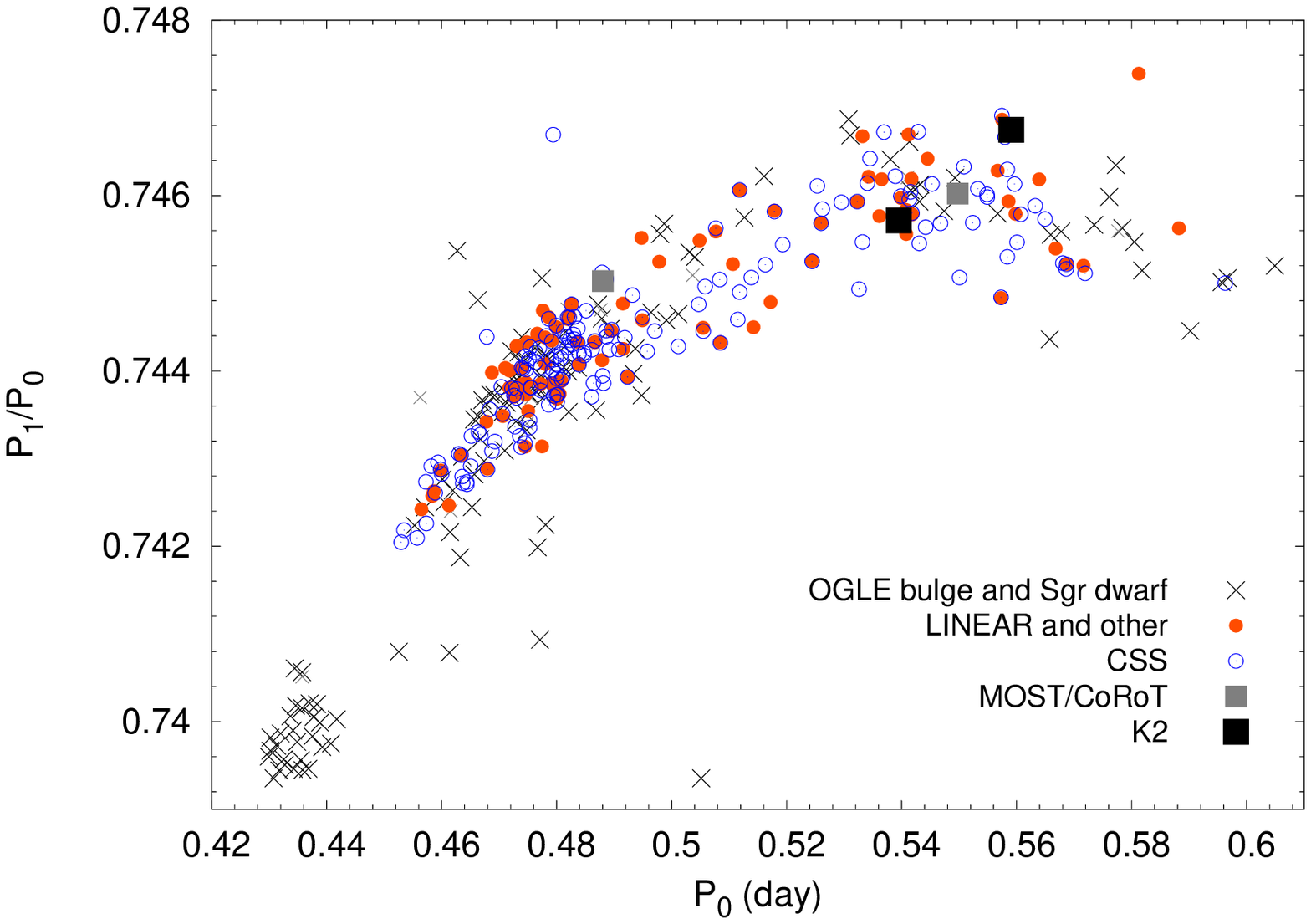}
\caption{Period ratio versus the period of the fundamental mode of the RRd stars, also known as the Petersen diagram. Small grey crosses: OGLE bulge and Sgr dwarf stars, their sequence continues to even smaller values. Black squares: the two K2 stars; orange dots: stars detected in the LINEAR, ASAS, NSVS and SuperWASP surveys; blue circles: our CSS sample from Appendix \ref{appx}; grey squares: CoRoT ID 0101368812 (left) and AQ Leo (right). Multiple identifications were not filtered. References for the data sources can be found in Section \ref{sec_hd_model}.} 
\label{petersen}
\end{figure}

The two K2 stars have similar pulsation properties. Their fundamental-mode periods are $P_0 = 0.53943$~d and $P_0 =  0.55900$~d, while the period ratios are $P_1/P_0 = 0.74581$ and $P_1/P_0 = 0.74678$ for EPIC 60018653 and 60018662, respectively. These values put them into the long-period end of the Petersen diagram that displays the period ratios against the fundamental-mode period of the stars (Figure \ref{petersen}). In fact, EPIC 60018662 is near to the long-period edge of the area covered by double-mode stars.

We constructed a Petersen diagram to put the two stars into broader context. The diagram in Figure \ref{petersen} includes the largest sample of field RRd stars constructed so far. One of us (RP) searched for bona-fide double-mode stars in the Catalina Surveys Catalog of Periodic Variable Stars \citep{cssfull} and identified 130 new RRd variables. The details of this search and the list of stars are presented in Appendix \ref{appx}. We also included the stars identified in the LINEAR \citep{poleski14}, ASAS/NSVS \citep*{wils05,nsvs,bw06, asas}, and SuperWASP \citep{wils10} databases. Stars from the Sagittarius dwarf galaxy and the OGLE bulge survey provide the background (grey crosses; \citealt{cseresnjes01,oglerrd}). It is interesting to note that even with the CSS sample, field stars outside the bulge seem to be restricted to $P_0 \geq 0.45$~d and $P_1/P_0 \geq 0.742$, whereas the Sgr and bulge population extends down to $P_0 \approx 0.35$ d and $P_1/P_0 \approx 0.726$ \citep{oglerrdmodels}.

The precise period values of pulsation modes in a star depend strongly on the fundamental parameters: mass, luminosity, effective temperature, and metallicity of the star. Therefore the simultaneous detection of two radial modes in one target can put strong constraints on its physical parameters. This is especially important for mass determination, as we currently have no other means to measure RR Lyrae masses. Although a few promising binary candidates were recently discovered by \citet{hajdu2015}, dynamical masses will take several years if not decades to measure. According to the calculations of \citet{szabo04}, RRd stars are confined to a narrow parameter range. We carried out a pilot study to see whether we can find good hydrodynamic models effectively. 

First we estimated the effective temperature, mass, luminosity and metal content ranges based on the pulsation model results of \citet{szabo04}. Compared to their Figure 6/a, the period parameters of the two stars fall well below the $Z=0.001$ models, therefore we selected two OPAL opacity tables with $Z=0.0001$ and 0.0003 metal contents. 

Then a grid of non-linear hydrodynamic models was computed with the Florida-Budapest turbulent convective code (details of the code are presented by \citet{kollath01} and \citet{kollath02}). Luminosity was varied between $L =48-60\, {\mathrm L_\odot}$, and two temperature and mass ranges were selected: one between $T_{eff} = 6480$--$6540$ K, $M=0.72$--$0.80\, {\mathrm M_\odot}$, and the other between $T_{eff} = 6410$--$6470$~K, $M=0.76$--$0.84\, {\mathrm M_\odot}$. Increments were 2 or 3 ${\mathrm L_\odot}$, 10 K and 0.2 ${\mathrm M_\odot}$. 

Models were iterated for 2000 fundamental-mode cycles and then the periods of both modes were calculated. We hand-picked 8 models close to the periods of EPIC 60018653 and 6 models close to 60018662 and iterated them for a further 2000 cycles. After that, 7 models converged slowly to single-mode pulsation and only 7 double-mode models remained. Six of those models are actually pairs that differ only in the temperature step. The results agree with the findings of \citet{szabo04} that non-linear RRd models exist only in narrow parameter ranges. 

\begin{figure}
\includegraphics[width=1.0\columnwidth]{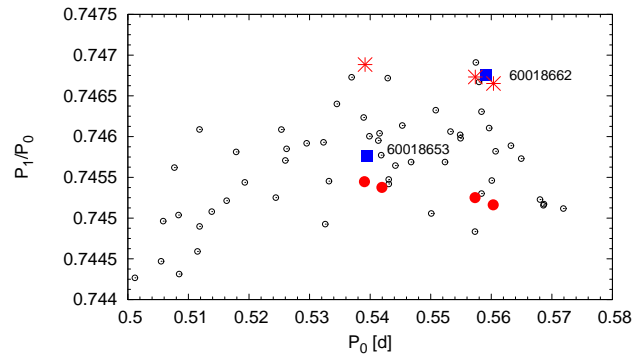}
\caption{Hydrodynamic model results compared to the two RRd stars (blue squares) on the Petersen diagram. The small circles are the CSS RRd stars from Appendix \ref{appx}. Large red dots are the $Z=0.0003$ models, large crosses are the $Z=0.0001$ models.} 
\label{rrd_models}
\end{figure}

As Figure \ref{rrd_models} illustrates we could not fit both stars with the same metal content value. For EPIC 60018653, two models at $T_{eff} = 6410$--$20$~K, $M=0.76\, {\mathrm M_\odot}$, $L =48\, {\mathrm L_\odot}$ and $Z=0.0003$ are relatively close to the star on the Petersen diagram. Two models fall very close to EPIC 60018662 in this parameter plane: $T_{eff} = 6430$--$40$ K, $M=0.78\, {\mathrm M_\odot}$, $L =52\, {\mathrm L_\odot}$ and $Z=0.0001$. We included, for comparison purposes, three more models that have similar fundamental-mode periods but different period ratios than the stars. Details of all seven models are included in Table \ref{model_table}.

The mass values appear to be quite high: pulsation masses for RRab stars in the original \textit{Kepler} field, for example, fell below $0.66\, {\mathrm M_\odot}$ \citep{nemec11}. Unfortunately, in the absence of proper dynamic masses, all mass estimates are model-dependent. Low-mass linear model calculations can be fitted to the whole range of the Petersen diagram \citep{oglerrdmodels}. However, non-linear RRd models in the Florida-Budapest code appear only at higher masses, typically above $0.7\, {\mathrm M_\odot}$ and they are still consistent with evolutionary models. We also note that the ability of current non-linear pulsation codes to properly model double-mode pulsation is still a matter of debate; see, e.g.\ \citet{kollath02} and \citet{sm08} for opposing views. We also note that the lack of the $f_X$ mode in the models does not invalidate our findings. If we assume that it is a non-radial mode, it could naturally arise in higher-dimension models with same the fundamental parameters.

We emphasize that these are just preliminary results from a pilot study. Calculating double-mode non-linear models is time-consuming, mainly because mode amplitudes may change very slowly. Nevertheless, we plan to carry out a more extensive survey after more RRd light curves will become available from the K2 mission. 

\begin{table}
\caption{Hydrodynamic model results for the two RRd stars, numbers correspond to EPIC 60018653 and 60018662. We also included the models where the first-overtone periods do not fit the observations, as indicated in Figure \ref{rrd_models}.} 
\begin{tabular}{lcccccc}
\hline
\noalign{\vskip 0.1cm}
Star      & Mass  & Lum. & $T_{eff}$ & $Z$    &  $P_0$   &  $P_1/P_0$   \\
~         & (${\mathrm M_\odot}$) & (${\mathrm L_\odot}$) & (K)  & -- & (d) & -- \\
\hline
\multicolumn{7}{c}{Good model fits}\\
653  &  0.76  &  48  &  6410  &  0.0003  &  0.54194  &  0.74538  \\
653  &  0.76  &  48  &  6420  &  0.0003  &  0.53906  &  0.74545  \\
662  &  0.78  &  52  &  6430  &  0.0001  &  0.56036  &  0.74665  \\
662  &  0.78  &  52  &  6440  &  0.0001  &  0.55737  &  0.74674  \\
\hline
\multicolumn{7}{c}{Models with different period ratios}\\
(653)  &  0.76  &  48  &  6410  &  0.0001  &  0.53917  &  0.74689  \\
(662)  &  0.78  &  51  &  6410  &  0.0003  &  0.56031  &  0.74516  \\
(662)  &  0.78  &  51  &  6420  &  0.0003  &  0.55731  &  0.74524  \\
\hline
\end{tabular}
\label{model_table}
\end{table}

\begin{figure*}
\includegraphics[width=1.0\textwidth]{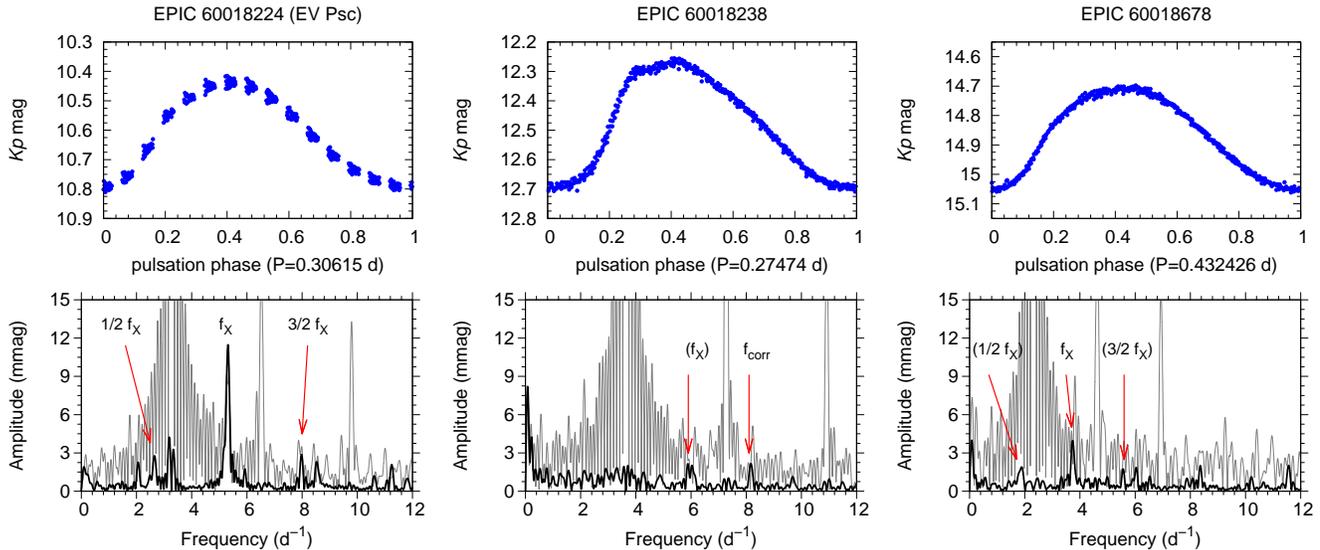}
\caption{Folded light curves and Fourier-spectra of the three non-modulated RRc stars. Original spectra are shown with thin grey lines; the black thick lines are the residuals after prewithening with the $nf_1$ $(n=1,2,\dots5)$ frequencies. The $f_X$ peak and the corresponding half-integer peaks are indicated. $f_{\rm corr}$ is the instrumental signal of the correction manoeuvres. Labels of low-significance peaks (S/N ratio between 3 and 4) are in parentheses. }
\label{rrc_folded}
\end{figure*}

\subsubsection{Additional modes in RRd stars}
The stars AQ Leo and CoRoT ID 0101368812 both showed additional modes that could not be explained by linear combinations of the fundamental mode and the first overtone. Comparison with the Fourier-parameters we derived from the K2 stars revealed striking similarities to these double-mode stars. One particular mode that falls to the frequency ratio of $f_1/f_X \approx 0.615$, was identified in both stars, and it has been recently identified in a modulated RRd star from the OGLE database \citep{smolecrrd} as well. Figure \ref{rrd} shows that we detected the same additional mode at frequency ratios $f_1/f_X = 0.6162$ in EPIC 60018653 and $f_1/f_X = 0.6166$ in 60018662. The same mode was reported not only in RRd stars but in RRc stars as well (Section \ref{sect_rrc}).

We detected the subharmonics, or half-integer peaks of the $f_X$ mode ($1/2$ and $3/2\,f_X$) in EPIC 60018653. The presence of these frequencies indicates that the mode experiences period doubling. The peaks are broadened and their centres are slightly off from the exact half-integer values at $0.499\,f_X$ and $1.495\,f_X$, respectively. Temporal variations in period doubling created forests of peaks in the spectra of RRab stars observed in the nominal mission and the highest-amplitude peaks were also displaced from the exact values \citep{pd}. Temporal variations have been detected in the $f_X$ mode and its subharmonics in the \textit{Kepler} RRc sample \citep{moskalik14}. It is very likely that the broadened peaks in the K2-E2 sample are caused by temporal variations in the period doubling of the $f_X$ mode as well. The $1/2\, f_X$ peak is marginally detectable in EPIC 60018662 too, but only with a signal-to-noise ratio of 2.5. 

\citet{aqleo} observed the same $1/2\, f_X$ peak in AQ Leo, but identified this peak tentatively as the parent frequency, $f_i$, and the true $f_X$ additional mode as $2\,f_i = f_{ii}$. \citet{chadid} reported two other additional modes ($f_3$, $f_4$) in the \textit{CoRoT} star. However, the frequency of their $f_4-f_1$ combination peak is equal to $1.49 f_X$. If we accept this peak as a subharmonic, then $f_4$ itself becomes a combination peak ($1/2\, f_X - f_1)$ instead of an independent mode, making the star more similar to the rest of the RRc sample. Similar combinations can be detected in EPIC 60018653 too. We can conclude that not only the $f_X$ additional mode, but its period doubling effect is observable in all four RRd stars measured by space photometric missions so far. 

\subsection{First-overtone stars}
\label{sect_rrc}
Four RRc stars were identified in the K2-E2 data, the same number as observed in the original \textit{Kepler} field. Three of those stars are normal first-overtone pulsators. The fourth star, CSS J235742.1-015022 is very different from the others: it exhibits strong amplitude and phase variations.

The three non-modulated stars have different pulsation periods and light-curve shapes, as illustrated in Figure \ref{rrc_folded}. EPIC 60018238 has the shortest period ($P=0.27474$~d) and a very pronounced hump feature just before maximum light. We detected two different additional modes with very small amplitudes. The star seems to exhibit the ubiquitous $f_X$ mode, although we detected two, closely-spaced peaks at $f_1/f_X = 0.6154 $ and $f_1/f_X' = 0.6048$ instead of a single frequency, with low significances (S/N ratios are $\sim 3.3$). We also identified the drift-correction frequency at $ 8.16\, {\mathrm d^{-1}}$ in the star.

The light curves of the other two RRc stars, EV Psc (EPIC 60018224) and EPIC 60018678 show almost no shock-related features (humps or bumps). The pulsation period of EV Psc exhibits strong beating with the 1766-second long-cadence sampling period, so it is not covered continuously (Figure \ref{rrc_folded}). We detected the $f_X$ mode in both stars at frequency ratios of $f_1/f_X = 0.6146$ (EV PSc) and 0.620 (60018678). The additional mode is very pronounced in EV Psc: the scatter of the folded light curve in Figure \ref{rrc_folded} is caused primarily by the mode itself. Moreover, we also detected the half-integer frequencies (1/2 and $3/2\,f_X$) in EV Psc, corresponding to period doubling in the $f_X$ mode. Half-integer peaks can be located but are marginal in EPIC 60018678. It seems that the (period-doubled) $f_X$ mode is intimately connected to the first radial overtone: it was observed not only in the K2 RRc stars but in the original \textit{Kepler} sample, in two \textit{CoRoT} stars, and the RRd stars described above as well \citep{moskalik14,szabo14}. 

In fact the $f_X$ mode has been detected in first-overtone Cepheids first, from the data collected by the OGLE surveys \citep{mk08,moskalik_rrc}. More recently, the $f_X$ mode has been identified in 147 RRc stars in the OGLE bulge sample and in 14 RRc and 4 RRd stars in the globular cluster Messier 3, respectively, further confirming its ubiquity \citep{netzel15,jurcsik15}. Yet the origin of the mode is still unclear: \citet{dziem} hypothesised that high-degree \textit{f}-modes ($\ell=42-50$) could explain the period ratios but those would require quite high intrinsic mode amplitudes. 

\subsubsection{Variation of the additional modes}
\begin{figure}
\includegraphics[width=1.0\columnwidth]{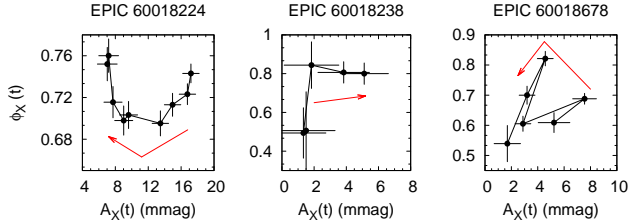}
\caption{Time variation of the amplitude and phase of the $f_X$ in the non-modulated RRc stars (also known as loop diagrams). Arrows denote the direction of progression. Variation can be detected in all three stars: EPIC 60018224 (EV Psc) shows particularly strong decrease in the amplitude of $f_X$. }
\label{f2}
\end{figure}

Beside the double peak in EPIC 60018238, we observed residual power in the Fourier spectra of both EPIC 60018224 and 60018678 after prewithening with $f_X$. These findings indicate that the $f_X$ mode undergoes amplitude and/or phase variations, similar to the findings of \citet{moskalik14} and \citet{szabo14} for \textit{Kepler} and \textit{CoRoT} stars, respectively.  We split the data sets into 1.5-day (EPIC 60018238 and 60018678) and 1-day (EV Psc) bins and calculated the Fourier-parameters in the bins. As Figure \ref{f2} illustrates, all stars show some variation, even on this short time scale. It is of course possible that there are actually multiple unresolved, separate peaks in the spectra but the considerably longer observations in the prime mission showed no signs of closely-spaced peaks in other RRc stars.

We carried out the same analysis on the RRd stars as well, but we did not detect significant amplitude variations in the $f_X$ mode. However, we found the uncertainties of the amplitude and phase data to be much higher than in the RRc stars, possibly because of the more complicated frequency content. Variation in the $f_X$ mode of the star CoRoT ID 0101368812 was detected by \citet{szabo14}, therefore we expect that the science data of K2 will be sufficient to probe the variation of these modes in other RRc and RRd stars.

\subsection{A strongly modulated RRc star}
The fourth star, CSS J235742.1-015022 is located within the subframe of an engineering target star, EPIC 60042292, but it has no EPIC identification itself. The K2 observations revealed that it shows very strong amplitude and phase modulation, as seen in Figure \ref{rrc_mod_lc}. The Blazhko effect exists in first-overtone RR Lyrae stars but it is less common than in their fundamental-mode siblings. \citet{mizerski} found a 10 per cent incidence rate in the OGLE-II bulge data while \citet{nagykovacs} derived 7.5 per cent for the Large Magellanic Cloud based on MACHO data, although both numbers should be considered as lower limits only. Recent surveys, such as the Northern Sky Variability Survey and the All-Sky Automated Survey identified several modulated field RRc stars \citep{nsvs,asas}. The Blazhko Stars in the Galactic Field database currently lists 55 modulated RRc stars \citep{blasgalf}. First-overtone stars with obvious Blazhko effect have not been observed before from space, though we note that \citet{szabo14} found a very small side peak in one of the \textit{CoRoT} RRc stars (ID~0105036241) that might be connected to amplitude and/or phase variations.

\subsubsection{Modulation properties of the RRc star}
\begin{figure}
\includegraphics[width=1.0\columnwidth]{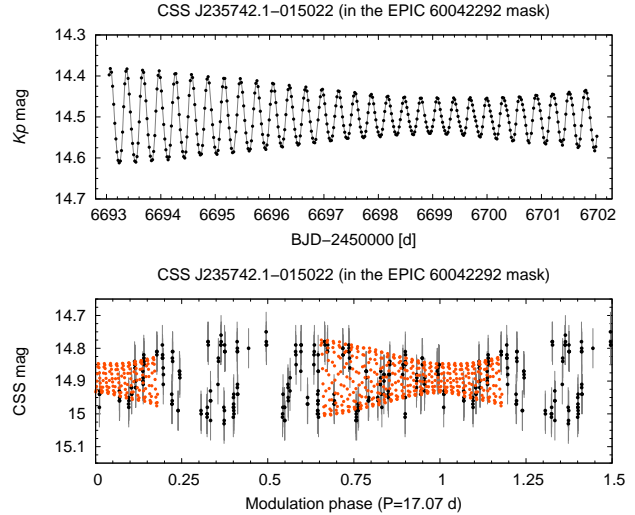}
\caption{Top: K2-E2 light curve of the modulated RRc star. Bottom: the Catalina Sky Survey (black) and K2 data (orange, shifted to the CSS brightness level), and folded with the modulation period. }
\label{rrc_mod_lc}
\end{figure}

The K2-E2 data covers less than a modulation cycle, therefore we analysed the CSS observations to determine the modulation period. We identified a single potential side peak, $f_1+f_m$, and determined a tentative modulation period of $P_m=17.07$~d. Comparison with the K2-E2 data revealed that this is indeed the correct value and it has been stable for more than 8 years, over the duration between the start of CSS and the K2 observations (lower panel of Figure \ref{rrc_mod_lc}). The detection of one modulated star out of eight RRc targets in the current \textit{Kepler} and K2 sample, agrees with the findings of previous studies and confirms the lower occurrence of the Blazhko effect among first-overtone stars. Adding the two \textit{CoRoT} stars, the ratio of modulated stars falls somewhere between 10--20 per cent, depending on whether we consider CoRoT ID~0105036241 modulated or not. More observations will be required of course to determine the occurrence rate more precisely. K2 is perfectly suited to provide the answer: we expect it to observe approximately 50-100 galactic RRc stars over the course of the mission. 

\begin{figure}
\includegraphics[width=1.0\columnwidth]{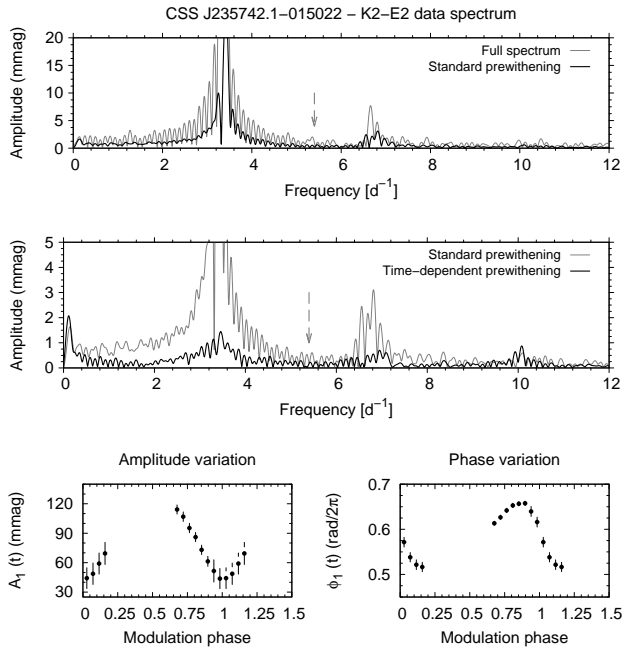}
\caption{Top: Fourier-spectra of the modulated RRc star, CSS J23542.1-015022 (within the mask of EPIC 60042292), original with thin grey line, prewithened with thick black line. Middle: residual spectra after standard (thin grey) and time-dependent (thick black) prewithening with the pulsation frequencies. The dashed arrows marks the expected position of the $f_X$ mode. Bottom: temporal variation of the amplitude and phase of the $f_1$ frequency, folded with the modulation period.}
\label{rrc_mod_sp}
\end{figure}

We divided the data of CSS J23542.1-015022 into twelve 0.75-day long segments and calculated the amplitudes and phases of the $f_1$ frequency peak in each bin to investigate the modulation properties of the star. The results are shown in the lower panels of Figure \ref{rrc_mod_sp}. The pulsation amplitude changes rather symmetrically, but the descending branch of the phase variation is much steeper than the ascending branch. In a phase-amplitude diagram, the modulation progresses in the clockwise direction, opposite the direction that was observed in another modulated RRc star, TV Boo \citep{tvboo}.

\subsubsection{Search for additional modes}

The Fourier spectrum of the star is shown in the upper and middle panels of Figure \ref{rrc_mod_sp}. We found no signs of additional frequencies in the star. As the K2-E2 data is shorter than the modulation period, the modulation sidepeaks cannot be resolved with conventional prewithening (constant Fourier parameters) properly. We applied an alternative, time-dependent prewithening method that subtracts non-stationary signals from the data, developed by \citet{moskalik14}. While this method resulted in much lower residuals, we still could not identify additional modes down to 0.4 millimag amplitude. Interestingly, ground-based observations of two other modulated RRc stars did not reveal any additional modes either, down to about 5 and 2 millimagnitudes for LS Her and TV Boo, respectively \citep{lsher,tvboo}. 

The $f_X$ mode seem to be a common occurrence in normal first-overtone stars that show only low-level temporal variability but not the Blazhko effect \citep{moskalik14}. The potential lack of the $f_X$ mode in strongly modulated RRc stars may raise several questions about mode excitation and the physical origin of the modulation. The K2 mission is ideally suited to expand our understanding about these problems. Further observations can also provide clues to whether the modulation in RRab and RRc stars is the same mechanism or is caused by different physical processes.

\begin{figure*}
\includegraphics[width=1.0\textwidth]{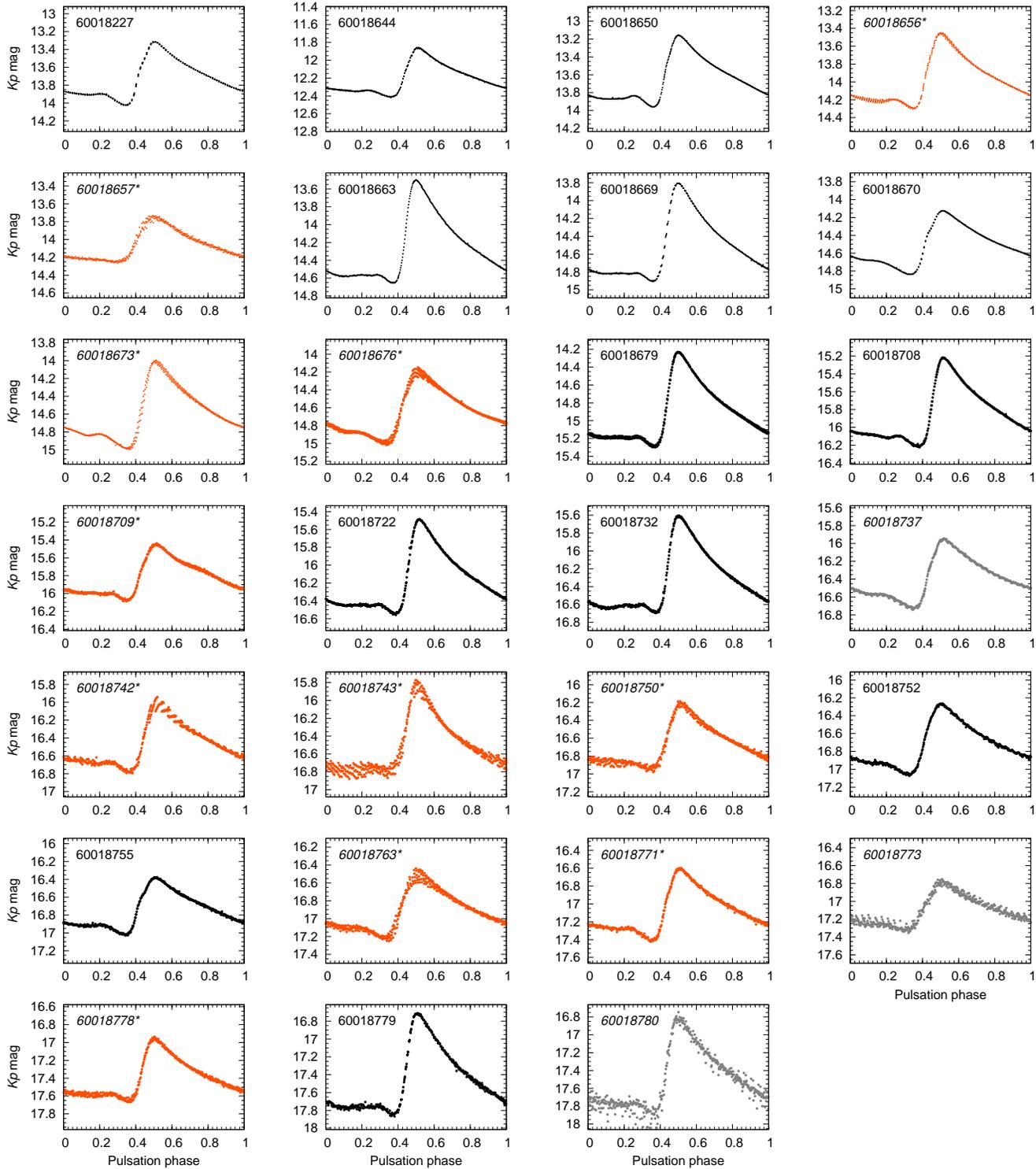}
\caption{Folded light curves of the 27 RRab stars in the K2-E2 sample, based on the periods presented in Table \ref{rrab_target_table}. In black, roman labels: non-modulated stars; in orange, and marked with asterisks: Blazhko stars; in grey, italic labels: uncertain. }
\label{rrab_folded}
\end{figure*}

\begin{table*}
\caption{Fundamental-mode RR Lyrae stars observed during the Two-Wheel Concept Engineering Test run of the K2 mission. Photometric ${\rm [Fe/H]}$ indices are included as well. The average uncertainty of the indices is $\pm 0.1$ dex.}
\begin{tabular}{lcccclcccc}
\hline
\noalign{\vskip 0.1cm}
EPIC ID     &   RA     &    Dec    &   \textit{Kp} &  Period & Name        &   ${\rm [Fe/H]}$  &    CSS ID No. & $P_m$ & Addtl.  \\
                 &  (deg)    &   (deg)   &    (mag)      &      (d)    &                    &                          &                         & (d)     & modes  \\
\hline
\noalign{\vskip 0.1cm}
\multicolumn{8}{c}{(a) Non-modulated RRab stars}\\
\noalign{\vskip 0.2cm}
60018227   &   0.83750     &   +3.39833    &   13.720   &   0.579041   &   ASAS   J000321+0323.9    &   $-$1.30   &   -- &   --  &   --   \\
60018644   &   354.15513   &   $-$2.21173   &   12.199   &   0.645043   &   ASAS   J233637$-$0212.7    &   $-$1.88   &   1001126012320 &   --  &   \textit{g}-mode?   \\
60018650   &   356.29939   &   $-$5.20608   &   13.636   &   0.630289   &   NSVS   11906749          &   $-$1.95   &   1004127008648 &   --  &   --   \\
60018663   &   355.36115   &   $-$5.54157   &   14.221   &   0.468800   &   NSVS   11904882          &   $-$1.34   &   1004127002457 &   --  &   --   \\
60018669   &   351.61480   &   $-$2.97015   &   14.490   &   0.465721   &   CSS   J232627.6$-$025814   &   $-$1.25   &   1004126052487 &   --  &   --   \\
60018670   &   357.53961   &   +1.02409    &   14.501   &   0.669217   &   NSV   26151              &   $-$1.29   &   1101128018965 &   --  &   --   \\
60018679   &   1.1001   &   $-$6.85657   &   14.888   &   0.494723   &   CSS   J000424.0$-$065123   &   $-$1.30   &   1007001029423 &   --  &   --   \\
60018708   &   356.48300   &   $-$7.47260   &   15.813   &   0.590650   &   CSS   J234555.9$-$072821   &   $-$1.63   &   1007126018552 &   --  &   --   \\
60018722   &   359.57559   &   $-$8.71836   &   16.133   &   0.482625   &   CSS   J235818.1$-$084306   &   $-$1.29   &   1009127050514 &   --  &   --   \\
60018732   &   352.76588   &   +1.06081    &   16.292   &   0.498292   &   CSS   J233103.7+010339   &   $-$1.56   &   1101126021705 &   --  &   --   \\
60018752   &   352.46439   &   +1.65528    &   16.710   &   0.594034   &   CSS   J232951.4+013919   &   $-$1.26   &   1101126033293 &   --  & (PD?)  \\
60018755   &   1.94723     &   +3.26733    &   16.740   &   0.681966   &   CSS   J000747.3+031602   &   $-$1.78   &   1104001007772 &   --  & -- \\
60018779   &   351.77498   &   $-$2.31793   &   17.415   &   0.465741   &   CSS   J232705.9$-$021904   &   $-$1.33   &   1001126009779 &   --  & --  \\
\noalign{\vskip   0.2cm}                                             
\multicolumn{8}{c}{(b) Modulated RRab stars}\\                                          
\noalign{\vskip   0.2cm}                                             
60018656   &   355.98041   &   $-$4.05090   &   13.956   &   0.565113   &   CSS   J234355.2$-$040303   &   $-$1.54   &   1004127030869 &   --  &   --   \\
60018657   &   353.86545   &   $-$5.72775   &   14.054   &   0.543388   &   CSS   J233527.6$-$054340   &   $-$1.38   &   1007125047274 &   84.3  & PD, F2  \\
60018673   &   357.75112   &   +4.87943   &   14.561   &   0.620309   &   CSS   J235100.2+045245   &   $-$1.49   &   1104128039542 &   --  & F2  \\
60018676   &   353.11519   &   $-$5.39302   &   14.633   &   0.699815   &   CSS   J233227.6$-$052334   &   $-$1.68   &   1004126005022 &   65.5  &   --   \\
60018709   &   352.8671   &   $-$2.62876   &   15.815   &   0.563396   &   CSS   J233128.1$-$023743   &   $-$1.19   &   1001126004067 &   --  &   --   \\
60018742   &   1.90229   &   $-$8.24558   &   16.460   &   0.594456   &   CSS   J000736.5$-$081444   &   $-$1.50   &   1007001004138 &   --  &   --   \\
60018743   &   352.99328   &   $-$4.92403   &   16.481   &   0.493351   &   CSS   J233158.3$-$045526   &   $-$1.75   &   1004126013659 &   --  & PD  \\
60018750   &   356.47678   &   $-$7.05360   &   16.658   &   0.548491   &   CSS   J234554.4$-$070312   &   $-$1.36   &   1007126026371 &   --  &   --   \\
60018763   &   357.49761   &   +4.62398   &   16.900   &   0.586196   &   CSS   J234959.4+043726   &   $-$1.19   &   1104128034480 &   --  &   --   \\
60018771   &   357.55888   &   +4.24208   &   17.065   &   0.553989   &   CSS   J235014.1+041431   &   $-$1.22   &   1104128027645 &   --  &   --   \\
60018778   &   355.94061   &   +1.86808   &   17.376   &   0.620342   &   CSS   J234345.7+015205   &   $-$2.00   &   1101127037513 &   --  &   --   \\
\noalign{\vskip   0.2cm}                                             
\multicolumn{8}{c}{(c) Possibly modulated RRab stars}\\                                          
\noalign{\vskip   0.2cm}  
60018737   &   355.31566   &   $-$4.21736   &   16.376   &   0.635342   &   CSS   J234115.7$-$041302   &   $-$1.26   &   1004127027477 &   --  &   --   \\
60018773   &   359.53266   &   +3.27389    &   17.094   &   0.595290   &   CSS   J235807.8+031626   &   $-$1.30   &   1104128008817 &   --  &   --   \\
60018780   &   1.56091     &   $-$4.50438   &   17.462   &   0.505453   &   CSS   J000614.6$-$043015   &   $-$1.08   &   1004001019454 &   --  & --  \\
\hline
\end{tabular}
\label{rrab_target_table}
\end{table*}

\section{Fundamental-mode stars}
We found 27 RRab-type stars in the K2-E2 data. Identifications and folded light curves are presented in Table \ref{rrab_target_table} and Figure \ref{rrab_folded}. The brightness range of the stars are similar to the original \textit{Kepler} RR Lyrae sample (excluding RR Lyr itself), between $Kp = 12.199$ and 17.462 magnitudes. We were able to extract high-precision photometry for almost all stars. The image of EPIC 60018780 is affected by dark CCD columns (negative video crosstalk). We extracted the flux from the remaining pixels, but because the pointing drifts affect the dark columns and the star differently, small jumps can be seen in the data, especially in the first part of the observations. Given the limited resolution of \textit{Kepler}, some stars are affected by blending. We see the strongest contaminations from another nearby object in EPIC 60018773 and 60018737. A more detailed evaluation and possibly image subtraction methods will be required to extract better photometry for these stars.

\subsection{The Blazhko stars}
Although the 8.9-day long time base is rather short, we identified amplitude variations in several stars. 11 out of the 27 stars show significant residual power in their Fourier spectra after the removal of the pulsation frequency and its harmonics. While the data are too short to resolve the modulation side peaks, the light curves show smoothly changing pulsation amplitudes. Therefore it is safe to assume that these stars are modulated. In 13 cases we found no signs of amplitude and/or phase change either in the spectra or in the amplitude variation data. Analysis of 3 stars led to inconclusive results. Therefore at least 40.7 per cent of the RRab stars seem to be modulated, but the number can be as high as $\sim 50$ per cent, in accordance with previous ground-based and \textit{Kepler} findings \citep{kbs,nemec13}. This is an impressive result if we consider that Blazhko periods are usually much longer than 9 days (in the \textit{Kepler} sample they ranged from 27 days to two years). 

In three cases, data problems prevented us from confirming or excluding the presence of modulation. EPIC 60018737 and 60018773 suffer from blending. In both cases, the Fourier spectra did not reveal any clear and significant modulation sidelobes. In fact, slight temporal variation can be detected in the amplitude of the $f_1$ peak, but the light curve itself is distorted and includes additional variation too. EPIC 60018780 is severely impacted from the negative video crosstalk which resulted in jumps in the light curve and elevated noise level in its Fourier spectrum.

\subsubsection{Modulation properties of RRab stars}
The K2-E2 data in themselves are too short to determine the modulation periods and amplitudes of the stars. While some of the brighter non-modulated stars were covered by multiple ground-based surveys, for the Blazhko stars we had to rely on the CSS data alone. The Catalina data has very sparse sampling in this area with 2-300 data points scattered over 7-8 years. Each field is covered 3-4 times per night and fields repeat after one or two weeks. This kind of sampling is well suited for asteroid hunting but less useful to determine the Blazhko periods for RR Lyrae stars. The precision of individual CSS data points is 0.05 mag for bright targets ($\sim 13$ mag) but reaches 0.1 magnitudes for targets fainter than 17 mag that is well below the (sub)mmag accuracy of the K2 data.

\begin{figure}
\includegraphics[width=1.0\columnwidth]{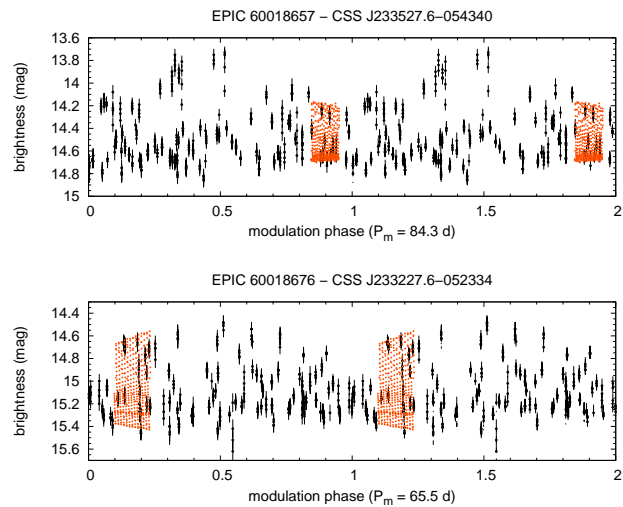}
\caption{Folded Catalina Sky Survey and K2-E2 light curves of two modulated RRab stars where we were able to estimate the Blazhko periods. }
\label{css_bla}
\end{figure}

We searched for possible modulation sidepeaks in the Fourier spectra of the CSS data. In some cases the beating between the pulsation period and the sampling resulted in modulation-like variations. To exclude them, we compared the modulation phases between the CSS and K2 light curves.  We were able to determine the Blazhko periods for two stars, EPIC 60018657 ($P_m = 84.3 \pm 0.3$ d)  and 60018676 ($P_m = 65.5 \pm 0.3$ d). Despite the poor coverage, the final light curves in Figure \ref{css_bla} line up quite well. Interestingly, this method failed for two stars that show strong amplitude variations, EPIC 60018742 and 60018743. We suspect that these stars have shorter modulation periods than the two we presented above and the CSS observations are simply too sparse in time to identify those periods properly.

Finally, EPIC 60018709 shows variations in its light curve shape and pulsation amplitude. After prewithening with the pulsation frequency and its harmonics, however, very little evidence can be found for side peaks anywhere but the $3f_1$ frequency value. Examination of the CSS data did not reveal any significant modulation side peaks, but the folded light curve indicates some amplitude variation around maximum light. Therefore we classified this star as a modulated RRab variable. 

\subsubsection{Additional modes in RRab stars}
\label{sec_rrab_addtl}
One of the major discoveries of \textit{Kepler}, regarding fundamental-mode RR Lyrae stars, was the identification of low-amplitude additional modes \citep{benko10,benko14}. With the help of model calculations, we were able to trace back the origin of period doubling to one of these modes, the ninth overtone \citep{pd,kmsz11}. We note in passing that although the ninth overtone itself has very low amplitude and is undetectable in the K2--E2 sample, period doubling itself can be still observed since it is not a classical beating between modes but a purely dynamical phenomenon. The strong resonance itself destabilizes the limit cycle of the fundamental mode that bifurcates into a new, period-doubled limit cycle that takes two pulsation periods to repeat. This in turn creates a series of half-integer peaks in the Fourier spectrum but those do not correspond to actual pulsation modes themselves.

We searched thoroughly for additional modes in the K2-E2 data set. For classical variables, the widely used criterion of accepting a peak is S/N $> 4$. In the long data sets available from \textit{CoRoT} and the \textit{Kepler} prime mission, however, combinations with $f_0$ were always observed for the detected additional peaks. Therefore, we also considered a peak (or a series of peaks) reliable and of stellar origin when it formed combination peaks with the fundamental mode, e.g., a regular pattern was detectable in the spectrum even if the combinations had a S/N ratio between 2 and 4. Four authors searched for periods independently and the results were compared to each other: we only accepted peaks that were detected by at least two authors. The results are included in Table \ref{rrab_target_table} and we detail them below.

\begin{figure}
\includegraphics[width=1.0\columnwidth]{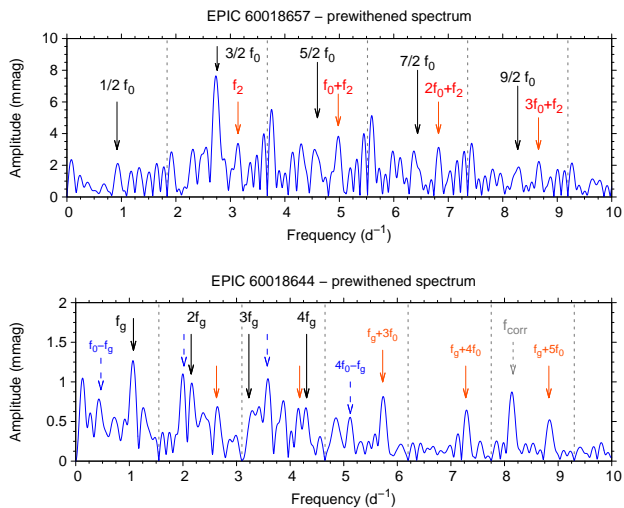}
\caption{Top: residual Fourier spectrum of the modulated star EPIC 60018657 after prewithening with the pulsation frequency and its harmonics (grey dashed lines). Identified or suspected additional peaks are labelled. In this case the Blazhko period is long (84.3 d), so the star shows little amplitude and phase variation over the K2 observations leading to very small modulation sidepeaks in the spectrum. Bottom: residual spectrum of EPIC 60018644, revealing the peaks corresponding to the $g$-mode suspected in the star. Thin solid and dashed arrows indicate the combination frequencies, $f_0+nf_g$ and $nf_0-f_g$, respectively. Some labels are not shown in the figure to avoid clutter. } 
\label{60018657}
\end{figure}

\begin{figure}
\includegraphics[width=1.0\columnwidth]{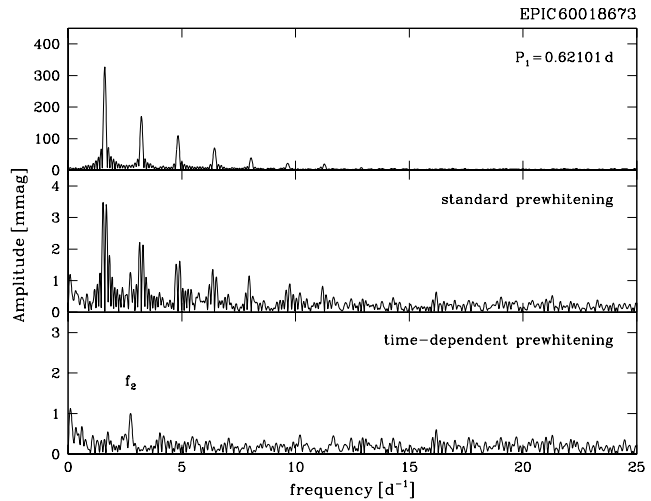}
\caption{Fourier spectrum of the RRab star EPIC 60018673 (top), and the residual spectra. Middle: residual of the standard prewithening with the main frequency and its harmonics: the strong, unresolved residual power hinders the detection of the additional modes. Bottom: residual of the time-dependent prewithening. The $f_2$ peak is unambiguously detected.}
\label{60018673}
\end{figure}

\begin{figure}
\includegraphics[width=1.0\columnwidth]{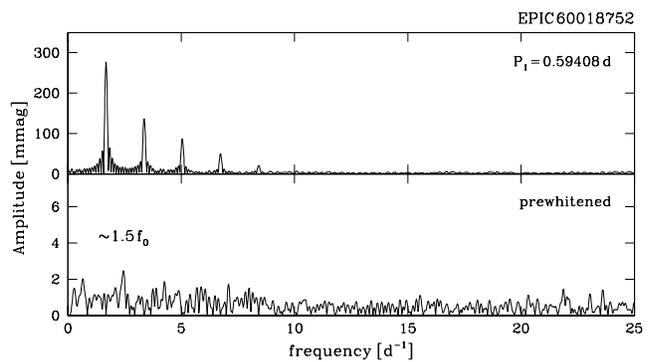}
\caption{Fourier spectrum of the RRab star EPIC 60018752 (top), and the residual spectrum (bottom). A single half-integer peak ($1.5f_0$) is tentatively detected, suggesting that period doubling might occur in the pulsation.}
\label{60018752}
\end{figure}

We identified the same instrumental frequency peak that we found in EPIC 60018238, $f_{\rm corr} \approx 8.16\, {\mathrm d^{-1}}$, in two RRab stars, EPIC 60018227 and 60018644, and excluded it from the analysis.

Despite our efforts, we found period doubling in only two stars, EPIC 60018657 and 60018743, and we have tentative detection in one more faint target (see Figure \ref{60018752}). We now know, however, that the amplitude of period doubling is very variable, often on short time scales. Long-term observations provided by \textit{CoRoT} and \textit{Kepler} can help the detection by accumulating multiple high-amplitude phases and lowering the noise in the Fourier-spectra \citep{benko14,szabo14}. On the other hand, short-term measurements, such as the K2-E2 can miss these signals entirely. 

The Fourier spectrum of EPIC 60018657 revealed not only half-integer peaks (($2n+1)/2\,f_0$), the tell-tale signs of period doubling, but other additional peaks as well. As the residual spectrum in Figure \ref{60018657} illustrates, the $f_2 = 3.14198$ d$^{-1}$ mode and its linear combinations with the $nf_0$ frequencies can be easily detected. Its frequency ratio, $f_0/f_2 = 0.5852$, is very similar to the ratios observed in other Blazhko stars in the \textit{Kepler} and \textit{CoRoT} samples \citep{benko14,szabo14}. This mode is hypothesized to be the second radial overtone, although it was not reproduced by non-linear hydrodynamic models so far. The reason of its absence could be that it is in fact a non-radial mode close to the frequency of the second overtone. It can also be hypothesized that the excited mode exists in a parameter space that 1D models have avoided so far, similar to the discovery of period doubling and the low-amplitude first overtone in some RRab models \citep{molnar12an}. We note, however, that the mode was not detected during the extensive survey of those models. Some excess amplitude still remains in the spectrum after the removal of the half-integer and $f_2$ peaks, but the dataset is too short to resolve it into further additional modes. We detected the $f_2$ peak firmly in one other star, EPIC 60018673, at $f_2=2.7380$ d$^{-1}$, with a period ratio of $f_0/f_2=0.5881$. This star shows stronger side peaks but we were able to detect $f_2$ with the time-dependent Fourier method, despite its small amplitude (1.1 mmag, Figure \ref{60018673}). 

Additionally, we identified an intriguing periodicity in EPIC 60018644. Although the peak at $f = 1.0774$ d$^{-1}$ itself has a signal-to-noise ratio of 3, we identified multiple significant peaks related to it (harmonics and combinations with $f_0$), therefore we accept this as a potential detection of a separate mode in the star. What makes this mode particularly interesting is that the period is longer than the fundamental mode itself, with a P/P$_0 = 0.695$ ratio, so it must be a \textit{g}-mode that are always non-radial. Similar modes have been identified recently in first-overtone stars observed by \textit{Kepler} and OGLE. The additional mode found in multiple OGLE stars has a period ratio of $\sim0.92-0.95$ (scaled to the expected period of the fundamental mode) that is very different from our value \citep{netzel15a}. On the other hand, \textit{Kepler} RRc stars show a multitude of modes between scaled period ratios from 0.38 to 0.97 \citep{moskalik14}. 

As the last column of Table \ref{rrab_target_table} indicates, we did not detect unambiguous additional $p$-modes in the non-modulated RRab stars, in accordance with previous results \citep{benko14,szabo14}, although a potential $g$-mode was identified in one star. Given the shortness of the data sets, however, both the tentative mode detections and the non-modulated status of the stars can be erroneous.

\subsection{Photometric ${\rm [Fe/H]}$ indices}

Light curve parameters, especially the $\phi_{31}$ Fourier phase relations of RRab stars ($\phi_{31}=\phi_3-3\,\phi_1$, as defined by \citealt{simonlee}) can be used to calculate photometric metallicity indices. \citet{nemec13} compared the Fourier parameters of the original \textit{Kepler} RR Lyrae sample with indices derived from spectra obtained with the CFHT and Keck telescopes. We applied the same relation (Eq.\ 3 from \citealt{nemec13}) to calculate the photometric ${\rm [Fe/H]}$ indices of the K2-E2 stars. The results are plotted in Figure \ref{metal} and included in Table \ref{rrab_target_table}. The K2-E2 sample is moderately metal-poor, with ${\rm [Fe/H]}$ values ranging from --1.08 to --2.00. Based on the experiences of \citet{nemec13}, we estimate the accuracy of the indices to be $\pm 0.1$ dex for non-modulated stars.

The accuracy is likely more limited for the modulated stars where the phase relation values change over the Blazhko cycle, but it is still acceptable. \citet{smolecfeperh} showed that filtering out the modulation side-peaks can lead to reasonably good [Fe/H] values. \citet{nemec13} also investigated the relation between the spectroscopic and photometric [Fe/H] determinations and found that the values are consistent for most Blazhko stars. Only three extremely modulated stars, less than 10 per cent of the overall sample, showed significant discrepancies. Based on the original \textit{Kepler} sample, we can estimate that about 90 per cent of the K2-E2 stars should closely follow the relation we used.

We consider the limited modulation phase coverage the strongest source of uncertainty in our study. Both the pulsation period and $\phi_{31}$ change during the modulation cycle: based on the values presented by \citet{nemec13}, we estimate the overall uncertainty of the [Fe/H] values for the Blazhko stars to be $\pm\, 0.15-0.2$ dex.  

\section{Conclusions and future prospects}
The \textit{Kepler} space telescope observed 1952 targets in Pisces for 8.9 days during the K2 Two-Wheel Concept Engineering Test run. We identified 32 RR Lyrae stars among the targets and one more as a background object in the target pixel mask of another star. We extracted the light curves from the target pixel files with the \textsc{PyKE} software tool \citep{pyke}. The analysis of the photometric data yielded the following results:

\begin{figure}
\includegraphics[width=1.0\columnwidth]{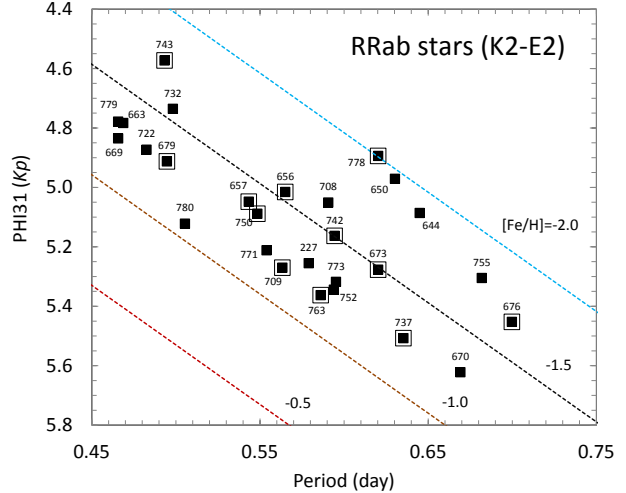}
\caption{Fundamental periods versus $\phi_{31}$ Fourier phase relations of the RRab stars. The relation of the two measures provide the photometric metallicities of the fundamental-mode stars that are included in Table \ref{rrab_target_table}. Modulated stars are indicated with additional boxes around the squares. Dashed lines correspond to various iso-metallicity lines, based on Eq.\ 3 of \citet{nemec13}.}
\label{metal}
\end{figure}

\begin{itemize}
\item We identified 2 double-mode (RRd), 4 first-overtone (RRc) and 27 fundamental-mode (RRab) stars in the field. Most of the stars were discovered by the Catalina Sky Survey \citep{cssfull}.
\item The $f_X$ additional mode was detected in the two RRd and the three non-modulated RRc stars, although with varying degrees of significance. The mode appears to be period-doubled and varies over time in three out of five stars. 
\item One of the RRc stars, CSS J125742.1--015022, turned out to be a modulated first-overtone star. This is the first time that a Blazhko-RRc star was observed by a space photometric mission. The modulation period of the star is 17.07~d and both the pulsation amplitude and phase show strong variations. It is the only RRc star where no additional modes were found. 
\item Indications of a potential $g$-mode was found in one RRab star, EPIC 60018644, at a period ratio of P$_0$/P$_g$ = 0.695.
\item Even from such short dataset, we found indications that at least 11 out of 27 RRab stars (41 per cent) may be modulated. We detected period doubling and the (presumed) second overtone in multiple stars. Based on the Fourier-parameter relations, the [Fe/H] values of the stars fall between $-1.08$ and $-2.00$ dex.
\item As a by-product of the target search for the K2 campaigns, we re-examined the `RRd' and `Blazhko' classifications of the Catalina Periodic Variable Star Catalog and identified 165 bona-fide double-mode stars throughout the sky, 130 of which were previously unknown. For a brief description, see Appendix \ref{appx}.
\end{itemize}

The results promise a bright future for the K2 mission. Even from these very short datasets, we were able to identify all new phenomena that space-based photometric missions discovered in RR Lyrae stars so far. The 75-day long campaign data will provide much better frequency resolution and refined light curve extraction methods will improve the photometric precision. 

Losing the ability of \textit{Kepler} to observe stars for several years was unfortunate and will be very hard to reproduce. The silver lining of K2 mission, however, is its flexibility to observe many different stars. This single field contained 33 RR Lyrae variables. Based on our estimates and target selection so far, this number of expected RR Lyrae stars will vary between approximately 10 and 200 for the forthcoming fields \citep{k2_ibvs}. With a sustained observing program, we will be able to answer new questions the previous missions could not: what is the true ratio of modulated RRc stars? Are all non-modulated RRab stars strictly periodic? Is the $f_X$ mode present in all RRd and RRc stars? Finally, K2 will be able to detect RR Lyrae stars in globular clusters, halo streams, and hopefully in nearby dwarf galaxies as well, broadening the scope of the mission even further. 

\section*{Acknowledgments}
This paper includes data collected by the \textit{Kepler} spacecraft during the K2 `Two-wheel Concept Engineering Test' operations. The authors gratefully acknowledge the \textit{Kepler} team, the Guest Observer Office, and Ball Aerospace, whose outstanding efforts have made these results possible. Funding for the \textit{Kepler} spacecraft is provided by the NASA Science Mission Directorate. This work has been supported by the ESA PECS Contract No. 4000110889/14/NL/NDe, and the Lend\"ulet-2009 and LP2014-17 Young Researcher Programs of the Hungarian Academy of Sciences. The research leading to these results has received funding from the European Community's Seventh Framework Programme (FP7/2007-2013) under grant agreements no.\ 269194 (IRSES/ASK), no.\ 312844 (SPACEINN), and no.\ 338251 (StellarAges). PM and RS are supported by the Polish National Science Center through grant DEC-2012/05/B/ST9/03932. CRTS and CSDR2 are supported by the U.S. National Science Foundation under grant AST-1313422. KK is grateful for the support of Marie Curie IOF grant 255267 SAS-RRL (FP7). The CSS survey is funded by the National Aeronautics and Space Administration under Grant No.\ NNG05GF22G issued through the Science Mission Directorate Near-Earth Objects Observations Program. This research has made use of the SIMBAD database, operated at CDS, Strasbourg, France.

\appendix

\section{Double-mode stars in the Catalina Sky Survey}
\label{appx}

\begin{figure}
\includegraphics[width=1.0\columnwidth]{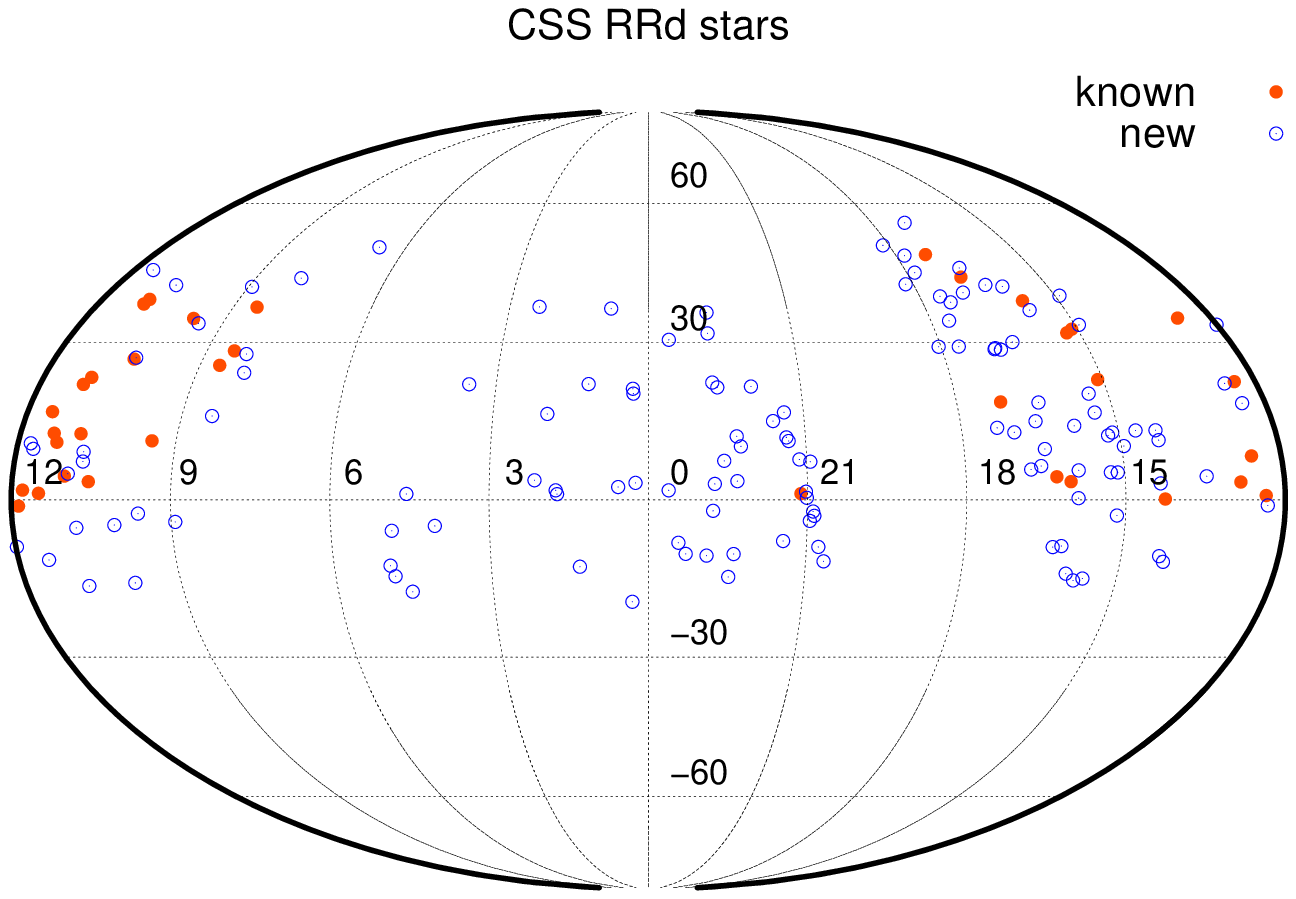}
\caption{Distribution of the CSS double-mode RR Lyrae stars in the sky, in Equatorial coordinates. Blue circles are the new discoveries, orange dots are the already known stars. }
\label{rrd_map}
\end{figure}

As a side-product to the K2 investigations we also identified 130 new RRd stars throughout the sky. These are very valuable targets, because their physical parameters can be estimated from hydrodynamic model calculations, as section \ref{sec_hd_model} illustrates. The Catalina Sky Survey provided us perhaps with the richest database of RR Lyrae stars for target selection for all K2 fields that are not too close to the Galactic plane. As usually only 1-5 RRd stars can be observed in a K2 campaign, proper identification is especially important. This work is also the continuation of the search for bona-fide RRd stars in the LINEAR sample by \citet{poleski14}. The stars we found are distributed between declinations $-30^\circ$ and $60^\circ$, except for a $\sim30$-degree band along the Galactic plane (Figure \ref{rrd_map}).

​Double-mode RR Lyr stars in the Catalina sample were separated from other RR Lyr by \citet{cssfull}, based on the visual examination of the light curves. No prewhitening was performed, thus, such classification can be uncertain as indicated by the misclassification of the two stars in the K2-E2 field. As the rare RRd stars are valuable targets, we searched for bona-fide double-mode pulsators among the stars that \citet{cssfull} classified as `RRd' or `Blazhko' stars. The photometric data were downloaded from \texttt{http://catalinadata.org/} \citep{drake09}. We found 461 stars classified as RRd and 70 stars classified as Blazhko for which at least 100 data points were obtained. Double mode pulsations were found in 161 and 5 stars, respectively. This means that only 35 per cent of the stars classified as `RRd' in Catalina catalogue are in fact double-mode pulsators. We checked that the fraction of confirmed RRd stars does not change if we limit to stars with at least 300 data points. Most of the stars that were erroneously identified as `RRd' in the CSS variable catalogue actually show Blazhko-type modulation. One star turned out to have two different CSS identifications (CSS\_J104003.0+414504 = CSS\_J104003.2+414503), bringing the total number of confirmed RRd stars to 165. We checked the literature for these stars and found that 35 of them were previously identified, most in the LINEAR survey data (29 stars, \citealt{poleski14}). Six stars were previously found by \citet{clementini00,bw06,asas,hw08}, and \citet{wils10}. The number of confirmed RRd stars is increased by 130. Table \ref{rrd_table} presents all confirmed RRd stars and Table \ref{rrd_table2} lists their pulsation parameters. Figure \ref{petersen} compares the positions of newly identified stars on Petersen diagram with RRd stars from various other sources.

We also checked CSS photometry for the two K2-E2 RRd stars. After removing outlying data points from Catalina data we were able to confirm double-mode pulsations in both stars. This illustrates that more double-mode stars are hidden in Catalina lists of the RRab and RRc stars. However, those lists are much longer, with tens of thousands and more than 5000 stars, respectively, and will require further work to classify.

\begin{table*}
\caption{Sample table of the confirmed double-mode RR Lyae stars in the Catalina Surveys Periodic Variable Stars Catalog. The list includes 165 stars, based on the examination of the light curves of the stars originally labelled as `RRd' or `Blazhko' in the Catalog, along with the two K2 stars. Full table is available in the online version. }
\begin{tabular}{lrrccc}
\hline
\noalign{\vskip 0.1cm}
CSS name     &   RA~~~    &    Dec~~~   &    Brightness & CSS ID & Other ID   \\
~    &   (deg)~~    &    (deg)~~   &   (mag) &   ~  & ~ \\
\hline
CSS\_J001420.8+031214 & $   3.586833 $ & $   +3.20389 $ & $ 17.45 $ & 1104002007409 &                 ---   \\
CSS\_J001724.9+200542 & $   4.353750 $ & $  +20.09506 $ & $ 16.64 $ & 1121002007726 &                 ---   \\
CSS\_J001812.9+210201 & $   4.554042 $ & $  +21.03375 $ & $ 14.54 $ & 1121002027610 &                 ---   \\
CSS\_J001836.2-191522 & $   4.651083 $ & $ -19.25619 $ & $ 16.84 $ & 1018002009928 &                 ---   \\
CSS\_J003359.4+022609 & $   8.497833 $ & $   +2.43583 $ & $ 15.87 $ & 1101004049971 &                 ---   \\
CSS\_J004804.2+365310 & $  12.017833 $ & $  +36.88625 $ & $ 16.60 $ & 1138004006108 &                 ---   \\
CSS\_J011032.5+215708 & $  17.635500 $ & $  +21.95233 $ & $ 16.13 $ & 1121006050243 &                 ---   \\
CSS\_J011824.3-123318 & $  19.601583 $ & $ -12.55525 $ & $ 17.72 $ & 1012007028731 &                 ---   \\
CSS\_J014305.3+010549 & $  25.772167 $ & $   +1.09708 $ & $ 17.04 $ & 1101010021310 &                 ---   \\
\dots & ~ & ~ & ~ & ~ & ~ \\
CSS\_J082622.9+282405 & $ 126.595625 $ & $  +28.40142 $ & $ 16.84 $ & 1129040004714 &      LINEAR 658512  \\
\dots & ~ & ~ & ~ & ~ & ~ \\
\hline
\end{tabular}
\label{rrd_table}
\end{table*}

\begin{table*}
\caption{Sample table of the pulsation parameters of the confirmed double-mode RR Lyae stars. FM: fundamental mode, O1: first overtone. Full table is available in the online version. }
\begin{tabular}{lccccccc}
\hline
\noalign{\vskip 0.1cm}
CSS name     &  FM period & FM period & FM ampl.\  & O1 period & O1 period & O1 ampl.\ & Per.\ ratio \\
~    &   (day)   &    uncert.  & (mag) & (day) &   uncert.  & (mag) & ($P_1/P_0$)\\
\hline
CSS\_J001420.8+031214 & $ 0.51930994 $ & $ 0.00000166 $ & $ 0.23 $ & $ 0.38711458 $ & $ 0.00000056 $ & $ 0.39 $ & $ 0.74544034 $ \\
CSS\_J001724.9+200542 & $ 0.48002554 $ & $ 0.00000174 $ & $ 0.29 $ & $ 0.35712249 $ & $ 0.00000074 $ & $ 0.41 $ & $ 0.74396560 $ \\
CSS\_J001812.9+210201 & $ 0.55837503 $ & $ 0.00000198 $ & $ 0.19 $ & $ 0.41615750 $ & $ 0.00000064 $ & $ 0.34 $ & $ 0.74530106 $ \\
CSS\_J001836.2-191522 & $ 0.46299846 $ & $ 0.00000198 $ & $ 0.28 $ & $ 0.34403253 $ & $ 0.00000067 $ & $ 0.40 $ & $ 0.74305329 $ \\
CSS\_J003359.4+022609 & $ 0.48368923 $ & $ 0.00000208 $ & $ 0.21 $ & $ 0.36009918 $ & $ 0.00000087 $ & $ 0.29 $ & $ 0.74448459 $ \\
CSS\_J004804.2+365310 & $ 0.46874750 $ & $ 0.00000070 $ & $ 0.38 $ & $ 0.34831989 $ & $ 0.00000079 $ & $ 0.37 $ & $ 0.74308640 $ \\
CSS\_J011032.5+215708 & $ 0.47302180 $ & $ 0.00000138 $ & $ 0.33 $ & $ 0.35178257 $ & $ 0.00000051 $ & $ 0.40 $ & $ 0.74369209 $ \\
CSS\_J011824.3-123318 & $ 0.52620846 $ & $ 0.00000320 $ & $ 0.27 $ & $ 0.39247062 $ & $ 0.00000147 $ & $ 0.38 $ & $ 0.74584628 $ \\
CSS\_J014305.3+010549 & $ 0.47555267 $ & $ 0.00000161 $ & $ 0.32 $ & $ 0.35372017 $ & $ 0.00000058 $ & $ 0.40 $ & $ 0.74380861 $ \\
\dots & ~ & ~ & ~ & ~ & ~ & ~ \\
CSS\_J082622.9+282405 & $ 0.47390930 $ & $ 0.00000138 $ & $ 0.22 $ & $ 0.35260694 $ & $ 0.00000044 $ & $ 0.35 $ & $ 0.74403887 $ \\
\dots & ~ & ~ & ~ & ~ & ~ & ~ \\
\hline
\end{tabular}
\label{rrd_table2}
\end{table*}

\end{document}